\documentclass[twocolumn,showpacs,floatfix,prl,superscriptaddress]{revtex4}

\usepackage{float}
\usepackage{graphicx}

\newcommand{\bk}{{\bf k}}

\newcommand{\bJ}{{\bf J}}
\newcommand{\tJ}{{\tilde J}}
\newcommand{\btJ}{{\bf \tilde J}}

\newcommand{\beq}{\begin{eqnarray}}
\newcommand{\eeq}{\end{eqnarray}}
\newcommand{\beqq}{\begin{eqnarray*}}
\newcommand{\eeqq}{\end{eqnarray*}}

\begin{document}

\title{Topological Crystalline Insulators and Dirac Octets in Anti-perovskites}

\author{Timothy H. Hsieh}
\email{thsieh@mit.edu}
\author{Junwei Liu}
\email{liujunweish@gmail.com}
\author{Liang Fu}
\email{liangfu@mit.edu}
\affiliation{Department of Physics, Massachusetts Institute of Technology, Cambridge, MA 02139}

\begin{abstract}
We predict a new class of topological crystalline insulators (TCI) in the anti-perovskite material family with the chemical formula A$_3$BX. Here the nontrivial topology arises from band inversion between two $J=3/2$ quartets, which is described by a generalized Dirac equation for a ``Dirac octet".  
Our work suggests that anti-perovskites are a promising new venue for exploring the cooperative interplay between band topology, crystal symmetry and electron correlation.

\end{abstract}

\maketitle

Topological crystalline insulators (TCIs)\cite{fu} are new topological phases of matter in two and three dimensions which exhibit metallic boundary states protected by crystal symmetry, unlike $Z_2$ topological insulators (TIs) that rely on time-reversal symmetry\cite{kane, zhang, moore}.  The first realization of topological crystalline insulators has recently been predicted\cite{hsieh} and observed in IV-VI semiconductors SnTe, Pb$_{1-x}$Sn$_x$Se and Pb$_{1-x}$Sn$_x$Te\cite{tanaka,dziawa,xu}. These TCIs exhibit a variety of novel topological electronic properties such as Dirac mass generation via ferroelectric distortion\cite{okada, serbyn} and strain-induced flat band superconductivity\cite{tang}, which are not only of fundamental interest but also may enable novel device applications\cite{liu, ezawa, fang, fzhang, qian}. On the theoretical frontier, the discovery of TCIs has sparked intensive efforts in classifying topological phases in different crystal symmetry classes\cite{mong, murakami, cfang, niu, slager, teo, chiu,  cxliu}. Given these developments, there is great interest in finding new TCI materials, especially those outside the family of narrow-gap semiconductors. Recent  proposals range from pyrochlore iridates\cite{fiete} and multilayer graphene\cite{kindermann} to heavy fermion compounds\cite{dai, sun}.

In this work, based on a combination of topological band theory, $k \cdot p$ model and first-principles calculations, we predict a new class of TCIs in the anti-perovskite material family A$_3$BX, with Sr$_3$PbO and  Ca$_3$PbO as two representatives. Here A denotes alkaline-earth or rare-earth metals (Ca, Sr, La), B denotes main group elements of the p-block (Pb, Sn), and X denotes nonmetals (C, N, O) \cite{widera}. The anti-perovskite structure is based on a perovskite but switches the positions of metal and nonmetal elements, see Fig.1a.  Anti-perovskites exhibit a wide range of interesting physical properties, such as superconductivity\cite{cava}, giant magnetoresistance\cite{tashiro}, negative thermal expansion\cite{takagi}, and magnetocaloric \cite{mc} effect, due to the cooperative interactions among lattice, spin, and charge degrees of freedom. Our prediction of TCIs in anti-perovskites thus opens up a promising new venue for topological phases in correlated electron systems.

\begin{figure}
\includegraphics[height=3in]{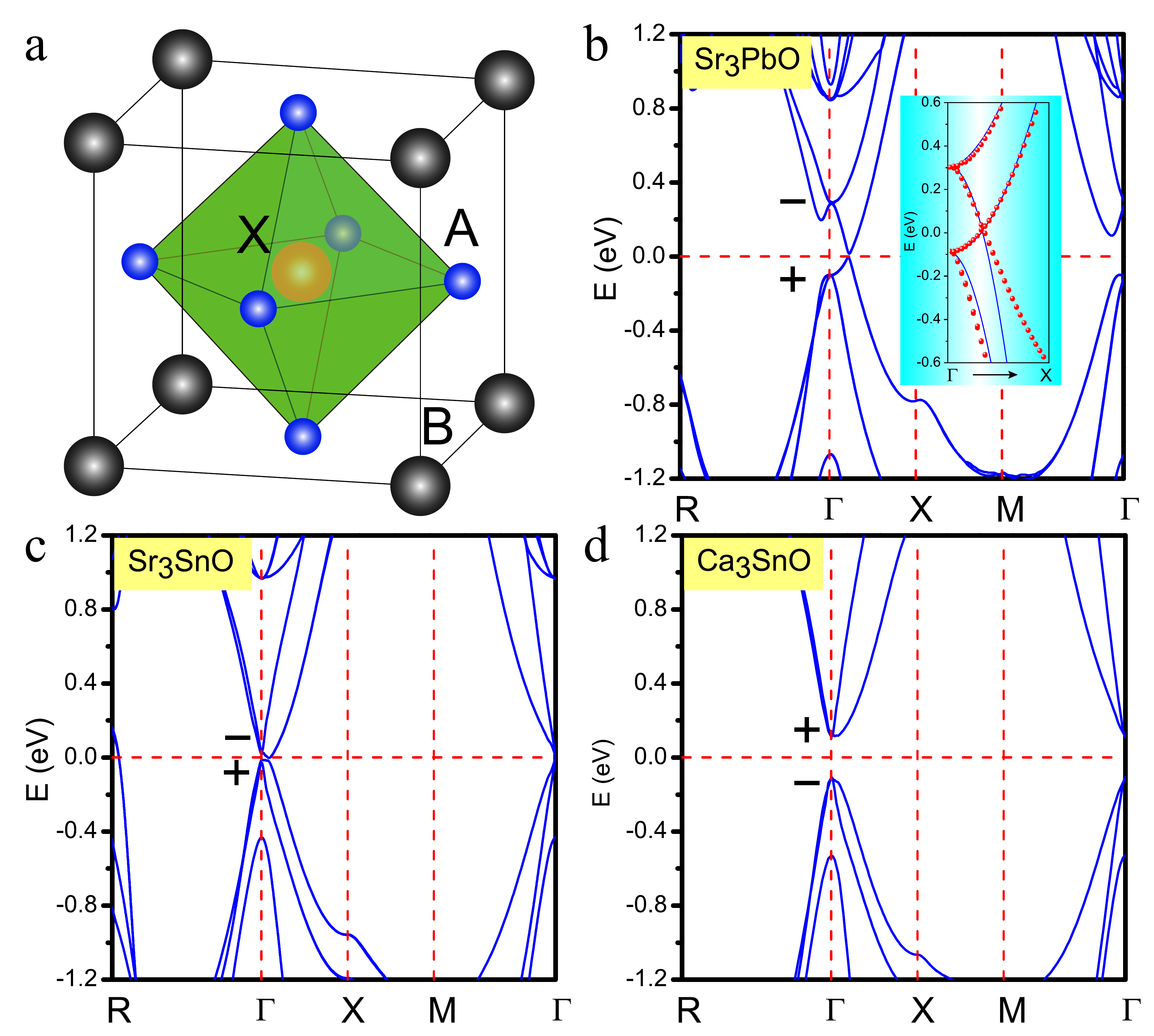}
\caption{(a) Crystal structure of the anti-perovskite $A_3 BX$. (b) Band structure of TCI Sr$_3$PbO in which the orbital character of valence and conduction bands within a Dirac octet is inverted ($\pm$ denote the parities of the band orbitals).  The inset depicts a fit with the $k \cdot p$ theory described in the main text. There is a small avoided crossing along $\Gamma X$ direction. (c) Sr$_3$SnO is near a topological phase transition with gap closing at $\Gamma$. (d) Ca$_3$SnO is a trivial insulator.  }
\end{figure}

Our work builds on recent theoretical calculations\cite{strain, ogata2} that noted an unusual low-energy band structure in Ca$_3$BiN and Ca$_3$PbO, in which both the conduction and valence bands at  the $\Gamma$ point are $J=\frac{3}{2}$ multiplets with four-fold degeneracy, which correspond to the $d$-orbitals of the A atom (Ca) and the $p$-orbitals of the B atom (Bi or Pb)
. These two sets of orbitals have opposite parities, leading to two possible band orderings at $\Gamma$. The normal ordering corresponds to the $d$-orbitals of A lying above the $p$-orbitals of $B$, which is expected for an ionic insulator (e.g., Ca$_3$SnO) made of A$^{2+}$, B$^{4-}$ and O$^{2-}$, hence topologically trivial. However, Ca$_3$BiN and Ca$_3$PbO were found to have an inverted band ordering in which the energies of the $d$- and $p$-orbitals are switched. Such a band inversion in anti-perovskites can be induced by decreasing the lattice constant or changing the chemical elements (e.g., Sn$\rightarrow$Pb). However, because of the four-fold degeneracy of the $J=\frac{3}{2}$ multiplet,
this band inversion does not change the product of parity eigenvalues of the valence bands. It then follows from the parity criterion\cite{parity} that anti-perovskites with inverted gaps are {\it not} topological insulators, as correctly pointed out in Ref.\cite{strain, ogata2}.

Here we show that despite being trivial in the $Z_2$ classification of TIs,  inverted anti-perovskites are TCIs in the same universality class as SnTe, which are protected by mirror symmetry and indexed by an integer topological invariant known as mirror Chern number\cite{teofukane}.
We find a nonzero mirror Chern number $|n_M|=1+1=2$ arises from the aforementioned band inversion between $J=\frac{3}{2}$ quartets in anti-perovskites. Remarkably, near the band inversion transition, the low-energy theory of anti-perovskites at $\Gamma$ is described by a novel generalization of three-dimensional Dirac fermion to eight-component and spin-3/2, which we term ``Dirac octet''.

We first present first-principles calculations of the band structures of three anti-perovskite compounds: Sr$_3$PbO, Sr$_3$SnO, and Ca$_3$SnO, see Fig.1b-d.
Our calculations were performed in the framework of density functional theory, by using the Perdew-Burke-Ernzerhof (PBE) generalized gradient
approximation\cite{PBE} and the projector augmented wave
potential\cite{PAW}, as implemented in the Vienna \emph{ab initio} simulation package \cite{VASP}. The energy cutoff of the plane-wave basis is 400 eV. The 11$\times$11$\times$11 Monkhorst-Pack \emph{k} points are used for bulk calculations. Structural relaxations are performed with forces converged to less than 0.001 eV/{\AA}, and spin-orbit coupling is included. To overcome the underestimation of band gap, we employed Heyd-Scuseria-Ernzerhof (HSE) screened Coulomb hybrid density functionals\cite{HSE} to calculate the bulk electronic structures.
By analyzing band parities at $\Gamma$ as mentioned above, we find that Ca$_3$SnO and Sr$_3$PbO belong to the normal and inverted regime respectively, with opposite orderings of $d$- and $p$-orbitals, while Sr$_3$SnO lies very close to the topological phase transition point.  Moreover, we find that Ca$_3$PbO, Ba$_3$PbO, Ca$_3$SiO, Ca$_3$GeO, Ca$_3$SnO, Ca$_3$BiN, and Sr$_3$BiN are also candidate TCIs (see Supplementary Material \cite{sm}).  

As a main result of this work, we now reveal the implication of the above band inversion in the context of TCI. For this purpose, we first develop $k\cdot p$ theory for this wide class of anti-perovskites.
We find that to linear order in $\bk$,  the cubic point group symmetry dictates the following eight-band $k \cdot p$ Hamiltonian describing the $J=\frac{3}{2}$ conduction and valence bands near $\Gamma$:
\beq
H(\bk)=m\tau_z + v_1 \tau_x \bk \cdot \bJ +v_2 \tau_x \bk \cdot \btJ \label{hk}
\eeq
Here ${\bf \tau}$ are Pauli matrices with $\tau_z=\pm 1$ labeling the valence and conduction band orbitals.  $\bJ$ are the spin-3/2 matrices and $\btJ$ are the only other set of $4$ by $4$ matrices which transform like the vector $\bk$ under the cubic point group\cite{sm}.  The form of the above $k\cdot p$ Hamiltonian is uniquely determined by requiring invariance under spatial inversion (represented by $P=\tau_z$), time reversal ($\Theta=e^{-i\pi J_y}K$, $K$ being complex conjugation), and discrete rotations of the cubic group $O_h$ generated by $\bJ$ which act on spin and spatial coordinates simultaneously.

The Hamiltonian (\ref{hk}) can be regarded as a novel generalization of Dirac Hamiltonian in three dimensions, involving an octet of spin-3/2 relativistic fermions that form two four-fold degenerate multiplets at $\bk=0$ protected by the cubic point group symmetries.
Moreover, Eq.(\ref{hk}) involves two velocities, leading to two sets of direction-dependent energy-momentum dispersions.
 In a special limit to be discussed later, Eq.(\ref{hk}) reduces to two identical copies of Dirac fermions.

We now analyze the mirror Chern number of $H(\bk)$. There are two sets of symmetry-equivalent mirror planes: (001) and (110).
Let us first consider symmetry of reflection with respect to the (100) mirror plane: $x \rightarrow -x$, which is represented by $M=PC_2=\tau_z e^{-i\pi J_x}$ where $C_2$ is rotation by $\pi$ about the x-axis.  Note that because $M^2=-1$, its eigenvalues are $\pm i$.
Due to this mirror symmetry, the eight-band Hamiltonian $H(k_x=0,k_y,k_z)$ commutes with $M$ and thus decouples into
two mirror subspaces, with mirror eigenvalue $\pm i$. The four states that span a given mirror subspace are given by the eigenstates of $J_x$, whose $\tau_z$ eigenvalue is locked to $j_x$ eigenvalue.
The corresponding four-band Hamiltonian $H_{\pm i}(k_y,k_z)$ within a mirror subspace is given by
\beq
H_{\pm i}(k_y,k_z)&=& \mp m( ie^{-i\pi J_x}) + \sum_{i=y, z} k_i (v_1 J_i + v_2 \tJ_i). \nonumber\\ \label{h2d}
\eeq

\begin{figure}
\includegraphics[height=2.5in]{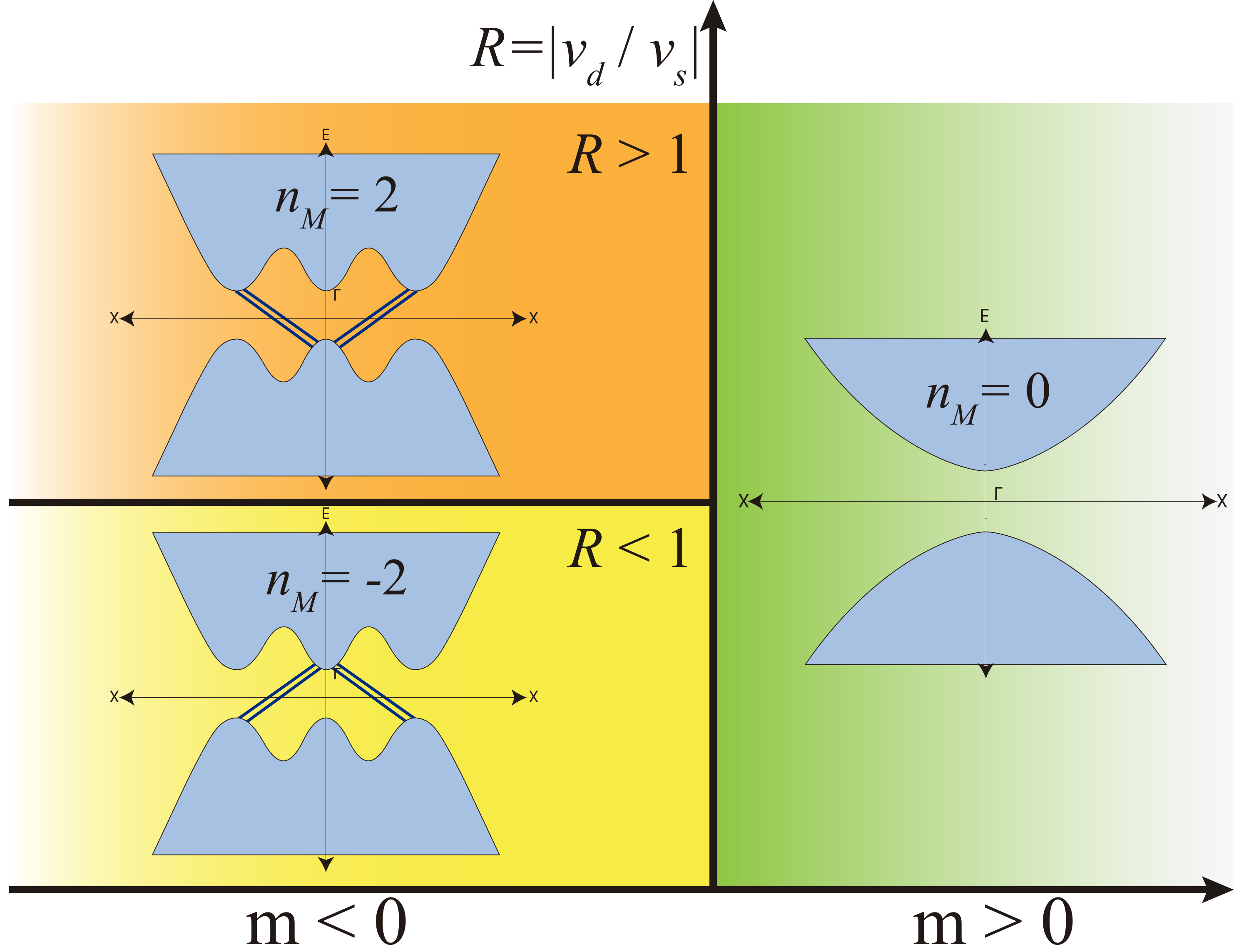}
\caption{Topological phase diagram for the minimal quadratic $k\cdot p$ theory describing a Dirac octet, for the (100) mirror plane.  The three phases have different mirror Chern number ($n_M$).  Each inset is a schematic band structure depicting bulk states in light blue and surface states in dark blue, cutting through the bulk gap.  The existence of the surface states is dictated by $n_M$.}
\end{figure}

Remarkably, we find the mirror Chern number $n_M$ depends on not only the sign of $m$ which controls the band inversion at $\Gamma$, but also the velocities  $v_1$ and $v_2$. We find it convenient to use the linear combinations $v_d\equiv v_1/2-v_2,v_s\equiv v_1 + v_2/2$ for velocities, and plot
the  topological phase diagram as a function of $m$ and $R\equiv | v_d/v_s| $ in Fig.2, which consists of three distinct phases with $n_M=0, 2$ and $-2$.
To obtain this result, we first consider a special limit $v_s=0$, for which the four-band Hamiltonian $H_{\pm i}(k_y,k_z)$ reduces to two identical flavors of two-component Dirac Hamiltonian:
\beq
H^{v_s=0}_{\pm i} (k_y,k_z) = \mp m \Gamma_0 + v_d (k_y \Gamma_1 - k_z  \Gamma_2).
\eeq
where the $4\times 4$ Gamma matrices are found to be $\Gamma_0 \equiv \sigma_z \otimes 1 $, $\Gamma_1 \equiv \sigma_x \otimes \sigma_x$ and $\Gamma_2 \equiv \sigma_y \otimes \sigma_x$ (written here in the $j_x$ basis), which forms the $SU(2)$ algebra.
When the Dirac mass $m$ changes sign, the Chern number of the $\pm i$ mirror sector changes by $\mp 2$, where the factor of two is due to the flavor degeneracy.
Therefore, as $m$ changes from positive to negative,  the mirror Chern number changes from $n_M=0$ to $n_M=2$. In accordance with convention,
we designate $m>0$ to  the trivial phase with normal band ordering as in Ca$_3$SnO, and $m<0$ to the TCI phase with inverted band ordering as in
Sr$_3$PbO.

To determine the mirror Chern number for $v_s \neq 0$, we need to account for potential gap closings even when $m$ is fixed to be nonzero.  For this, we go beyond linear order in $k$ and simply make the replacement
\beq
m \rightarrow m + \alpha k^2 
\eeq
where $\alpha>0$.  This is similar to the quadratic term in the BHZ model \cite{bhz}.  While several $O(k^2)$ terms are allowed by symmetry, we focus on the above for simplicity and find that it qualitatively reproduces the band dispersion from ab initio calculations (Fig.1b), especially in the inverted regime.

To gain intuition for the band dispersion, first consider the limit of zero hybridization between valence and conduction orbitals ($v_s=v_d=0$).  In the trivial regime ($m>0$), the conduction and valence bands do not cross.  However, in the inverted regime ($m<0$), the bands cross at $\sqrt{|m|/\alpha}$ and restoring the hybridization ($v_s,v_d$) opens a gap at this crossing.  

Importantly, we find that tuning the ratio $v_d/v_s$ closes and reopens this gap along the $\Gamma X$ directions, with criticality at $v_d/v_s=\pm 1$.  To model this gap behavior, we derive the most general $k\cdot p$ theory for this avoided crossing in the Supplementary Material and we state the result here:
\beq
H_{k_0}=m_0 \sigma_z +v_z \delta k_z (\gamma+s_z \sigma_x) + v (k_x s_y - k_y s_x) \sigma_x \nonumber
\eeq
Here $s_z=\pm 1$ denotes the Kramers pair ($j_z=\pm 3/2$), $\sigma_z=\pm 1$ denote valence and conduction band orbitals, $m_0$ is the mass term determining the gap, and $v$ and $v_z,\gamma$ govern the velocities of dispersion in the $x(y),z$ directions.  $\delta k_z \equiv k_z-k_0$, where $k_0$ is the momentum at which the gap is minimum, if $\gamma=0$.  Since there are four such $\Gamma X$ directions on the $k_x=0$ mirror plane, every time the four gaps close and reopen (corresponding to $m_0$ changing sign), the mirror Chern number (of the $k_x=0$ mirror plane) changes by 4.

Therefore, in going from the previously analyzed limit $|v_d/v_s|=\infty$ to $|v_d/v_s|=0$, the mirror Chern number changes from $n_M=2$ to $n_M=-2$ at $|v_d/v_s|=1$.  In the Supplementary Material, we provide a detailed derivation of this.  The phase diagram is shown in Fig. 2; we emphasize that in the inverted regime, the phases are always topologically nontrivial $|n_M|=2$.  Unlike the SnTe class of materials, in which band inversion of spin $1/2$ fermions at different points in momentum space add up to yield $|n_M|=2$, here the band inversion occurs at a single point $\Gamma$ and it is the spin $3/2$ nature of the octet which yields $|n_M|=2$.

Our theory thus captures and deduces the consequences of two essential features of the band structure of several anti-perovskites: 1) the band inversion of the octet at $\Gamma$ which as shown above gives rise to 2) a small gap (avoided crossing) at finite momentum along $\Gamma X$ in such band inverted anti-perovskites.  Previous works by Kariyado and Ogata \cite{kariyado1, ogata2} have focused on this finite $k$ avoided crossing at the Fermi energy; they attributed the smallness of the gap (15 meV for Ca$_3$PbO) to the combination of hybridization with orbital states away from the Fermi energy and spin-orbit coupling.  While prior works have emphasized the massive Dirac fermions at finite $k$ {\it near} $\Gamma$, the main feature of this work is the ``parent" Dirac octet {\it at} $\Gamma$, whose inverted nature gives birth to not only the Dirac fermions at finite $k$ but also the TCI phase.  

This nontrivial bulk topological invariant has the following consequences for surface states.  Consider any surface which respects reflection symmetry about the $(100)$ or equivalent mirror plane of the crystal.   Along the projection of the mirror plane to the surface, there will be two sets of gapless, counter-propagating surface states dictated by the mirror Chern number $n_M=\pm 2$ (see depiction in Fig. 2), similar to the case of SnTe\cite{hsieh}. The 
locking between mirror eigenvalue and directionality of these surface states depends on the sign of $n_M$. We note that a previous work by Kariyado and Ogata discovered surface states in a tight-binding model of of Ca$_3$PbO\cite{ogata2}, which are closely related to the TCI surface states.  However, the tight binding model they used produces a crossing, not avoided crossing, at finite $k$, and from this bulk gapless phase it is not possible to infer the band connectivity of the surface states or discuss their topological origin.  We leave a detailed study of the TCI surface states to future work.  

A similar analysis applies to the $(1\bar{1}0)$ and symmetry equivalent mirror planes (see Supplementary Material for derivation\cite{sm}), and we summarize the result here and in the phase diagram of Fig. 3: unlike the $(100)$ plane, the $(1\bar{1}0)$ has both trivial ($|n_M|=0$) and nontrivial ($|n_M|=2$) phases in the inverted regime ($m<0$).  Hence, we introduce the notation $(n_{M1},n_{M2})$ to capture the potentially different mirror Chern numbers of the $(100)$ and $(1\bar{1}0)$ planes. To fully determine these topological quantum numbers for each inverted anti-perovskite compound requires a careful analysis of first-principle results, which is left to future work. 

\begin{figure}
\includegraphics[height=2in]{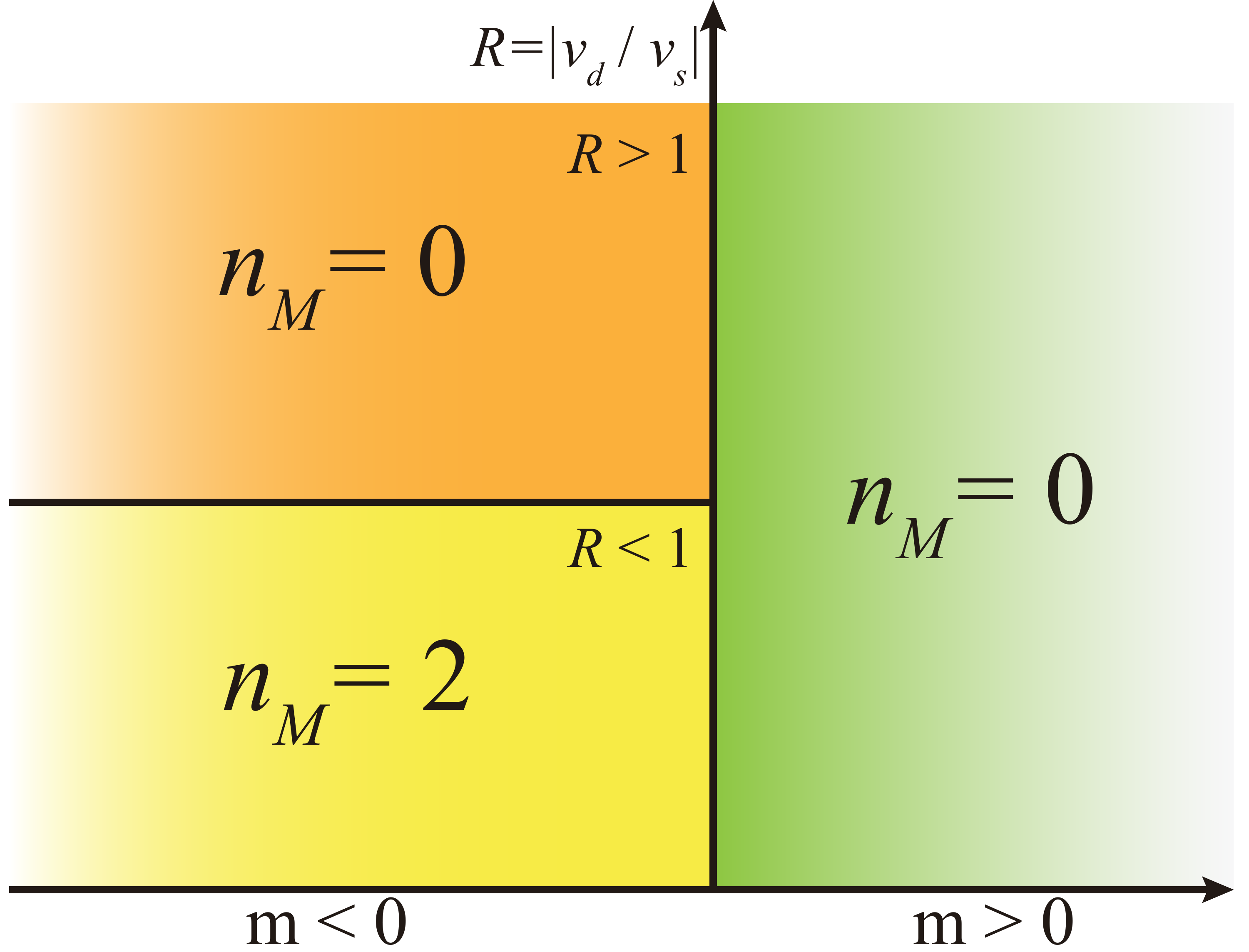}
\caption{Topological phase diagram for the minimal quadratic $k\cdot p$ theory of ($1\bar{1}0$) mirror plane.  The numbers $n_M$ are the mirror Chern number.}
\end{figure}

The above theory applies to many anti-perovskite materials such as Ca$_3$PbO and Sr$_3$PbO and captures both essential features of the band structure--an inverted octet at $\Gamma$ which gives rise to an avoided crossing at finite $k$.  As a result, we have demonstrated the existence of topological crystalline insulators in the anti-perovskite material class, with Ca$_3$PbO and Sr$_3$PbO as representative examples with mirror Chern number $|n_M=2|$ on the $(100)$ and symmetry related mirror planes.

The experimental feasibility of thin film and heterostructure growth, strain, and alloying may add to the already diverse features of this material class.  As thin films of SnTe are examples of two-dimensional topological crystalline insulators protected by mirror symmetry about the film plane \cite{liu}, it is possible that thin films of anti-perovskites may also be topologically nontrivial.  One may expect that both the inverted octet at $\Gamma$ and the small avoided crossing at finite $k$ can be tunable with layer thickness.  Moreover, given the wealth of phenomena in perovskite heterostructures and interfaces \cite{oxide, nagaosa, oxide2, oxide3}, it is likely that similarly diverse features can be found in the anti-perovskite counterparts.  Likewise, applying strain serves as yet another experimental knob on these two gaps which will affect the topological class of the material \cite{strain}.  Finally, by applying pressure or tuning the chemical composition to interpolate between a trivial and a nontrivial (TCI) anti-perovskite, as done in the case of Pb$_{1-x}$Sn$_x$Se \cite{dziawa}, one may be able to observe a bulk gap closing and topological phase transition.

We conclude that the anti-perovskite family hosts a rich variety of features, now encompassing topology and symmetry.  The existence of the Dirac octet at $\Gamma$ undergoing band inversion leads to their classification as topological crystalline insulators endowed with robust metallic surface states.  The presence of the high spin fermions fits naturally with the notion of TCIs, which allow for even number of band inversions unlike $Z_2$ topological insulators.  The Dirac octet fermions are quite distinct from typical four-component Dirac or two-component Weyl fermions, and the topological phase diagram for the full $k\cdot p$ theory to quadratic order is expected to be quite rich.  These new ingredients of topology, symmetry, and the high spin Dirac octet make the anti-perovskite family a promising playground for experimental and theoretical developments.

{\it Acknowledgement:}
This work is supported by NSF Graduate Research Fellowship No. 0645960 (TH) and DOE Office of Basic Energy Sciences, Division of Materials Sciences and Engineering under award DE-SC0010526 (LF and JL).

\section{Supplementary Material}

\subsection{4 by 4 Matrices Forming a Vector in the Cubic Point Group}
For the reader's convenience, we show the spin-3/2 matrices $\bJ$ here:
\beq
J_x&=&\left( \begin{array}{cccc}
0 & \frac{\sqrt{3}}{2} & 0 & 0 \\
\frac{\sqrt{3}}{2} & 0 & 1 & 0 \\
0 & 1 & 0 & \frac{\sqrt{3}}{2} \\
0 & 0 & \frac{\sqrt{3}}{2} & 0 \end{array} \right) \\
J_y&=&\left( \begin{array}{cccc}
0 & -i\frac{\sqrt{3}}{2} & 0 & 0 \\
i\frac{\sqrt{3}}{2} & 0 & -i & 0 \\
0 & i & 0 & -i\frac{\sqrt{3}}{2} \\
0 & 0 & i\frac{\sqrt{3}}{2} & 0 \end{array} \right) \\
J_z&=&\left( \begin{array}{cccc}
\frac{3}{2} & 0 & 0 & 0 \\
0 & \frac{1}{2} & 0 & 0 \\
0 & 0 & -\frac{1}{2} & 0 \\
0 & 0 & 0 & -\frac{3}{2} \end{array} \right)
\eeq

In addition to $\bJ$, the following matrices also transform as a vector under the cubic point group:
\beq
\tJ_x&=&\left( \begin{array}{cccc}
0 & \frac{\sqrt{3}}{4} & 0 & -\frac{5}{4} \\
\frac{\sqrt{3}}{4} & 0 & -\frac{3}{4} & 0 \\
0 & -\frac{3}{4} & 0 & \frac{\sqrt{3}}{4} \\
-\frac{5}{4} & 0 & \frac{\sqrt{3}}{4} & 0 \end{array} \right)\\~\nonumber\\
 \tJ_y&=&\left( \begin{array}{cccc}
0 & -i\frac{\sqrt{3}}{4} & 0 & -i\frac{5}{4} \\
i\frac{\sqrt{3}}{4} & 0 & i\frac{3}{4} & 0 \\
0 & -i\frac{3}{4} & 0 & -i\frac{\sqrt{3}}{4} \\
i\frac{5}{4} & 0 & i\frac{\sqrt{3}}{4} & 0 \end{array} \right)\\~\nonumber \\
\tJ_z&=&\left( \begin{array}{cccc}
-\frac{1}{2} & 0 & 0 & 0 \\
0 & \frac{3}{2} & 0 & 0 \\
0 & 0 & -\frac{3}{2} & 0 \\
0 & 0 & 0 & \frac{1}{2} \end{array} \right)
\eeq
After deriving these, we noticed that they had already been detailed in \cite{elder} and can also be understood as a linear combination of $\bJ$ and $\bJ^3$, which also transforms as a vector \cite{luttinger}.

\subsection{$k\cdot p$ Theory for Band Inversion at Finite $k$}

Here we derive the $k \cdot p$ theory describing the avoided crossing along the $\Gamma X$ direction, denoted as the $k_z$ direction in the following.  In \cite{ogata2}, a $k \cdot p$ theory was derived from a tight binding model.  Here, the form of the Hamiltonian is dictated by the following symmetries and their representations:
Inversion with time reversal ($is_y K\sigma_z$), fourfold rotation about the $z$-axis ($e^{is_z \pi/4}$), mirror symmetries about the $x,y$ axes ($is_x\sigma_z,is_y\sigma_z$), and mirror about $z$-axis followed by time reversal ($s_x \sigma_z K$) or inversion ($is_z$).  Here $s_z=\pm 1$ correspond to the Kramers pair ($j_y=\pm 3/2$) and $\sigma_z=\pm 1$ to the valence/conduction bands.  $K$ denotes complex conjugation.  The most general Hamiltonian allowed by these symmetries is
\beq
H&=&m_1 \sigma_z + m_2 s_z \sigma_x \nonumber \\
&+&k_z(w_0+w_1 \sigma_z + w_2 s_z \sigma_x) + v (k_x s_y - k_y s_x)\sigma_x \nonumber
\eeq
A unitary rotation generated by $s_z \sigma_y$ transforms the above into
\beq
H=m \sigma_z +v_z(k_z-k_0)(\gamma + s_z \sigma_x) + v (k_x s_y - k_y s_x)\sigma_x, \nonumber
\eeq
where $k_0$ is the location of the gap minimum if the ``tilt'' parameter $\gamma$ is zero, and $v,v_z$ are velocities.
Hence, we see that the gap is tuned by one parameter $m$, and when $m$ changes sign, the Chern number of the $k_x=0$ plane changes by 4 (there are four such avoided crossings on the plane).

\subsection{Detailed Analysis of (100) Mirror Plane}
Here we provide a detailed derivation of the phase diagram shown in Fig.2 of the main text.
To proceed, we find it convenient to use the following basis involving two sets of Pauli matrices for the four spin-3/2 states, denoted by $|\sigma_z=\pm 1,s_z=\pm 1\rangle$ and defined as follows:
\beq
|1, 1\rangle&=&-|3/2\rangle+|-3/2\rangle \nonumber \\
|1, -1\rangle&=&-|1/2\rangle+|-1/2\rangle \nonumber \\
|-1, 1\rangle&=&|3/2\rangle+|-3/2\rangle \nonumber \\
|-1, -1\rangle&=&|1/2\rangle+|-1/2\rangle,
\eeq
where the right hand sides are written in the $j_z$ basis.  
%

In this Pauli matrix basis, the Hamiltonian for the $\pm i$ mirror eigenvalue sectors of the (100) mirror plane is:
\beq
H_{\pm i} (k_y,k_z)&=& -m\sigma_z -k_z (v_s \sigma_x + v_d \sigma_x s_z) \nonumber \\
&+& k_y \left(\mp\frac{v_s}{2}\sigma_y
\pm v_d \sigma_y s_z-\frac{v_s\sqrt{3}}{2} \sigma_x s_y\right) \nonumber
\eeq


First consider the sector with mirror eigenvalue $+i$ and set the last term involving $s_y$ to zero.  At linear order in $k$, restoring the last term does not close the band gap and thus does not affect this analysis of the band topology.  In this case, $s_z$ is a conserved quantum number, and each subsector (labeled by $s_z=\pm 1$) reduces to a two-band Hamiltonian describing a massive Dirac fermion.  When the mass $m$ changes sign, the Chern number of each subsector changes by one.  If the orbital character of the valence and conduction bands is inverted ($m<0$) relative to an atomic insulator ($m>0$), then we can conclude that the norm of each subsector Chern number is $|n_{+i,s_z=\pm 1}|=1$.

This still leaves two scenarios, depending on $v_d,v_s$.  If $0.5<|v_d/v_s|<1$, then the $s_z=\pm 1$ sectors have opposite chirality and $n_{+i,s_z=\pm1}$ have opposite signs, suggesting that the total Chern number of the mirror eigenvalue $+i$ sector is $n_{+i}=0$.  Otherwise, $n_{+i,s_z=\pm1}$ have the same sign, and $|n_{+i}|=2$.  However, we now show that the first scenario is an artifact due to working to only linear order in $k$.

The essence of this artifact is the effect of the neglected $\sigma_x s_y$ term in the above analysis.  While this term does not close the band gap in the linear theory, it can and does affect the band gap in higher order theory, in which the fundamental band gap may be located at finite $k$.

In particular, in the extended $k \cdot p$ theory with $m\rightarrow m+\alpha k^2$, we find that tuning the $\sigma_x s_y$ term to zero closes the band gap for $0.5<|v_d/v_s|<1$ but does not close the gap for for either $0<|v_d/v_s|<0.5$ or $|v_d/v_s|>1$.  The latter two regimes are all we need to complete the analysis, however.  In both regimes, the Chern number changes by 2 as $m$ changes sign.  Figure 2 in the main text shows the only phase diagram consistent with these results involving changes in Chern number of 2 and 4 across vertical and horizontal phase boundaries, respectively.

\subsection{Analysis of $(1\bar{1}0)$ Mirror Plane}

Projecting onto the mirror eigenstates of the operator representing reflection about (1$\bar{1}$0) plane ($\tau_z e^{i\pi (J_x-J_y)/\sqrt{2}}$), we get the Hamiltonian
\beq
H_{+i}&=&-m\sigma_z - k_z \sigma_x (v_s + v_d s_z) \nonumber \\
&+&k_{xy}\sqrt{2}\left(\sigma_y (v_d-s_z \frac{v_s}{2})+\frac{1}{\sqrt{2}}\sigma_x(s_x-s_y) \frac{\sqrt{3}}{2}v_s\right) \nonumber
\eeq
where $k_{xy}\sqrt{2}$ is the distance along the (110) direction.  Once again, dropping the last term doesn't close the band gap at this linear order, and we find that when $m$ changes, the Chern number changes by 2 if $0.5<|v_d/v_s|<1$ and doesn't change otherwise.

However, adding the quadratic term $m\rightarrow m+\alpha k^2$ shows that the lack of change in Chern number is an artifact for $-0.5<v_d/v_s<0.5$; in this range, the neglected term closes the gap along the (110) direction.  In this extended $k\cdot p$ theory, there are again phase boundaries at $v_d/v_s=\pm 1$, where the gap at finite $k$ along (100) closes.  However, since there are only two such points on the (1$\bar{1}$0) plane, the Chern number only changes by 2.  Hence, in the inverted phase, we find Chern number 2 for $-1<v_d/v_s<1$ and zero otherwise, shown in the phase diagram in Fig. 3.

\subsection{Other TCI Candidates in the Anti-perovskite Material Class}
In Figures 4 and 5, we show the band structures of several more anti-perovskites which are TCI candidates.  These calculations were performed using DFT-PBE and HSE methods described in the main text.  However, for the materials shown in Fig. 5, which are near a topological phase transition at the $\Gamma$ point, we found that whether or not HSE is used affects the results; this dependence on method is consistent with \cite{zunger}.

\newpage

\begin{figure}
\includegraphics[height=1.7in]{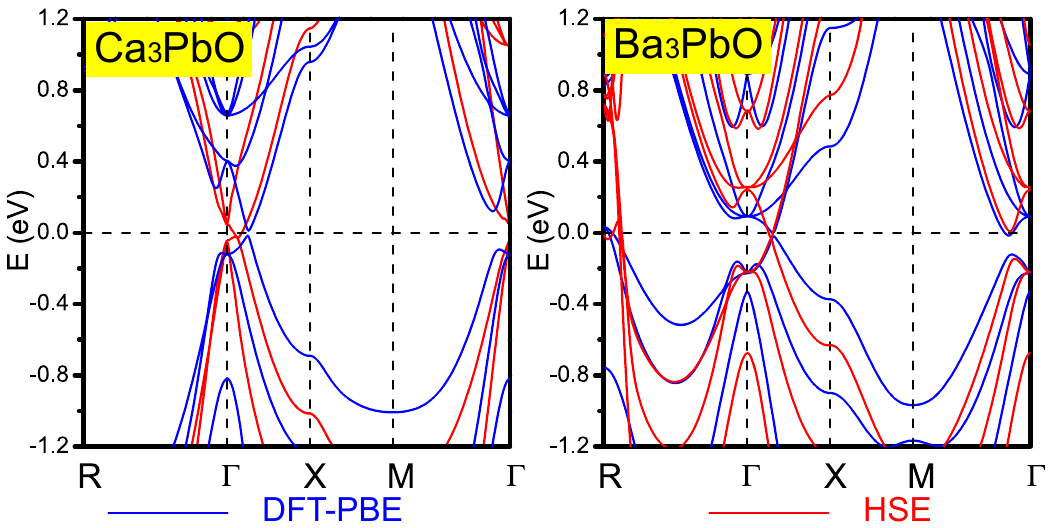}
\caption{Band structures of TCI candidates Ca$_3$PbO and Ba$_3$PbO, calculated using DFT-PBE method (blue) and HSE method (red).}
\end{figure}

\begin{figure}
\includegraphics[height=4.2in]{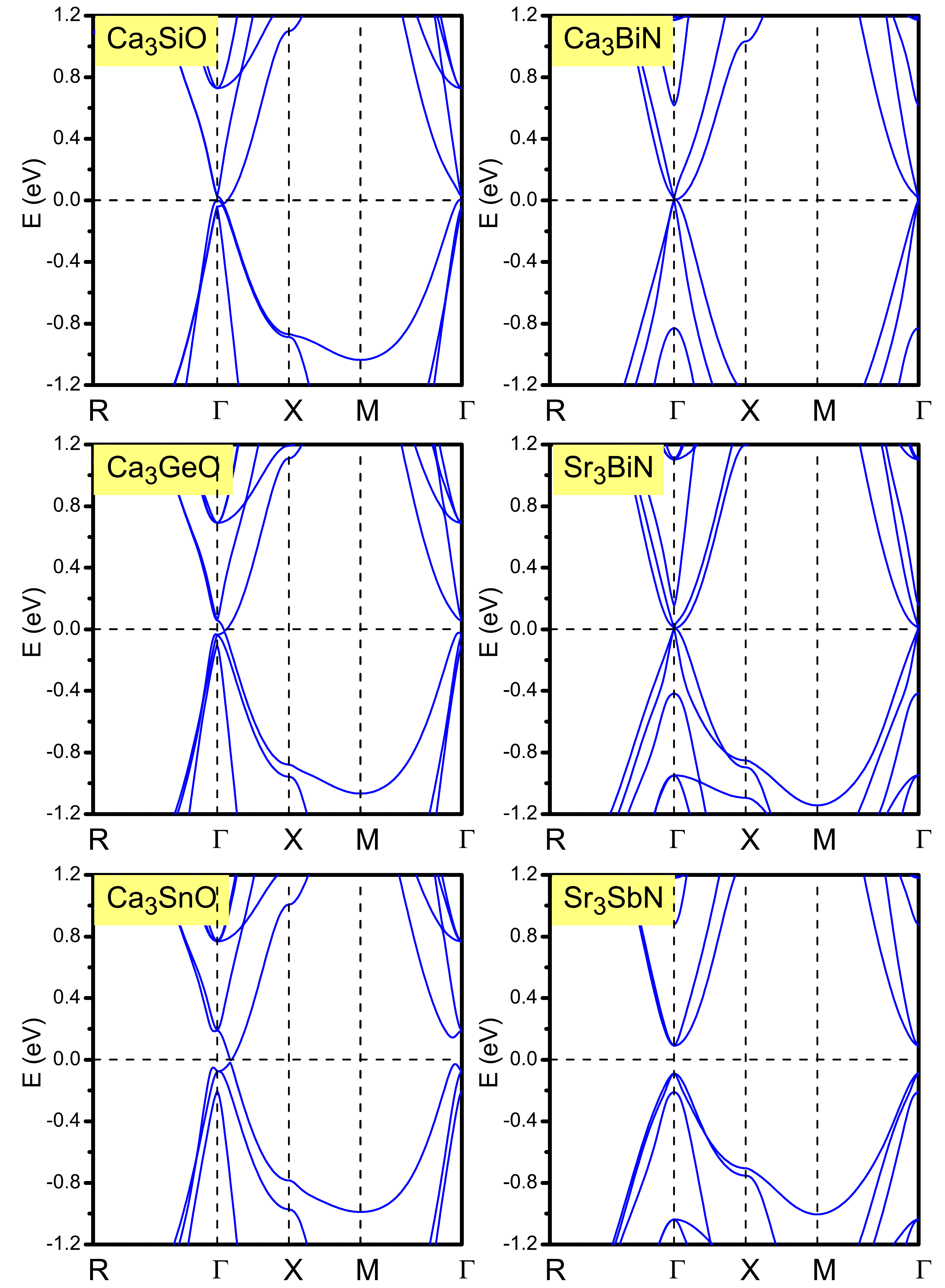}
\caption{Band structures of more TCI candidates, except for Sr$_3$SbN, which is topologically trivial and shown here for contrast.  These were calculated using DFT-PBE, and we found that using HSE affects the results.}
\end{figure}


\begin{thebibliography}{10}



\bibitem{fu}
L. Fu, Phys. Rev. Lett. {\bf 106}, 106802 (2011).

\bibitem{kane}
M. Z. Hasan and C. L. Kane, Rev. Mod. Phys {\bf 82}, 3045 (2010).

\bibitem{zhang}
X. L. Qi and S. C. Zhang, Rev. Mod. Phys. {\bf 83}, 1057 (2011).

\bibitem{moore}
J. E. Moore, Nature {\bf 464}, 194 (2010).

\bibitem{hsieh}
T. Hsieh, H. Lin, J. Liu, W. Duan, A. Bansil, and L. Fu. Nat. Comm. 3, 982 (2012).

\bibitem{tanaka}
Y. Tanaka, Z. Ren, T. Sato, K. Nakayama, S. Souma, T. Takahashi, K. Segawa, and Y. Ando. Nat. Phys. 8, 800 (2012).

\bibitem{dziawa}
P. Dziawa,	B. J. Kowalski,	K. Dybko,	R. Buczko, A. Szczerbakow, M. Szot, E. Lusakowska, T. Balasubramanian, B. M. Wojek, M. H. Berntsen, O. Tjernberg, and T. Story. Nat. Mat. 11, 1023 (2012).

\bibitem{xu}
S-Y. Xu, C. Liu,	N. Alidoust, M. Neupane, D. Qian, I. Belopolski,	J.D. Denlinger,	Y.J. Wang, H. Lin, L.A. Wray, G. Landolt, B. Slomski, J.H. Dil, A. Marcinkova, E. Morosan, Q. Gibson, R. Sankar,	F.C. Chou, R.J. Cava, A. Bansil, and M.Z. Hasan. Nat. Commun. 3, 1192 (2012).

\bibitem{okada}
Y. Okada, M. Serbyn, H. Lin, D. Walkup, W. Zhou,C. Dhital,M. Neupane,S. Xu,YJ Wang,R. Sankar,F Chou,A Bansil,4 M. Z Hasan,S.D. Wilson,L. Fu, and V. Madhavan. Science, 341, 1496 (2013).

\bibitem{serbyn}
M. Serbyn and L. Fu. Phys. Rev. B 90, 035402 (2014)

\bibitem{tang}
E. Tang and L. Fu. arXiv:1403.7523v1 (2014).

\bibitem{liu}
J. Liu, T. H. Hsieh, P. Wei,	W. Duan, J. Moodera, and L. Fu. Nat. Mat. 13, 178 (2014).

\bibitem{ezawa}
M. Ezawa. Phys. Rev. B 89, 195413 (2014).

\bibitem{fang}
C. Fang, M.J. Gilbert, and B.A. Bernevig. Phys. Rev. Lett. 112, 046801 (2014).

\bibitem{fzhang}
F. Zhang, X. Li, J. Feng, C. L. Kane, and E. J. Mele. arXiv:1309.7682 (2013).

\bibitem{qian}
X. Qian, L. Fu and J. Li. arXiv:1403.3952 (2014).

\bibitem{mong}
R. Mong, A.M. Essin, and J.E. Moore. Phys. Rev. B 81, 245209 (2010). 

\bibitem{murakami}
R. Takahashi and S. Murakami. Phys. Rev. Lett. 107, 166805 (2011).

\bibitem{cfang}
C. Fang, M.J. Gilbert, and B.A. Bernevig. Phys. Rev. Lett. 112, 106401 (2014).

\bibitem{niu}
P. Jadaun, D. Xiao, Q. Niu, and S.K. Banerjee. Phys. Rev. B 88, 085110 (2013).

\bibitem{slager}
R-J. Slager, A. Mesaros, V. Juricic, and J. Zaanen. Nature Physics 9, 98Ð102 (2013).

\bibitem{teo}
W.A. Benalcazar, J.C.Y. Teo, and T.L. Hughes. Phys. Rev. B 89, 224503 (2014).

\bibitem{chiu}
C-K Chiu, H. Yao, and S. Ryu. Phys. Rev. B 88, 075142 (2013).

\bibitem{cxliu}
C. X. Liu. arXiv:1304.6455 (2013).

\bibitem{fiete}
M. Kargarian and G.A. Fiete. Phys. Rev. Lett. 110, 156403 (2013).

\bibitem{kindermann}
M. Kindermann. arXiv:1309.1667 (2013).

\bibitem{dai}
H. Weng, J. Zhao, Z. Wang, Z. Fang, and X. Dai. Phys. Rev. Lett. 112, 016403 (2014).

\bibitem{sun}
M. Ye, J. W. Allen, and K. Sun. arXiv:1307.7191 (2013).

\bibitem{widera}
A. Widera and H. Schafer.  Mater. Res. Bull. 15 1805 (1980).

\bibitem{cava}
T. He, Q. Huang, A. P. Ramirez, Y. Wang, K. A. Regan, N. Rogado,
M. A. Hayward, M. K. Haas, J. S. Slusky, K. Inumara,
H. W. Zandbergen, N. P. Ong, and R. J. Cava. Nature 411, 54-56 (2001).

\bibitem{tashiro}
H. Tashiro. J. Kor. Phys. Soc. 63, 3, 299-301 (2013)

\bibitem{takagi}
K. Takenaka and H. Takagi. Appl. Phys. Lett. 87, 261902 (2005)

\bibitem{mc}
B. S. Wang, P. Tong,Y. P. Sun,X. B. Zhu, X. Luo, G. Li, W. H. Song, Z. R. Yang, and J. M. Dai J. Appl. Phys. 105, 083907 (2009).

\bibitem{strain}
Y. Sun, X-Q Chen, S. Yunoki, D. Li, and Y. Li Phys. Rev. Lett. 105, 216406 (2010).

\bibitem{ogata2}
T. Kariyado and M. Ogata. J. Phys. Soc. Jpn. 81 064701 (2012).

\bibitem{parity}
L. Fu and C.L. Kane, Phys. Rev. B {\bf 76}, 045302 (2007).

\bibitem{teofukane}
J. C. Y. Teo, L. Fu and C. L. Kane, Phys. Rev. B {\bf 78}, 045426 (2007).


\bibitem{PBE} Perdew, J. P., Burke, K. \& Ernzerhof, M. Phys. Rev. Lett. \textbf{77} 3865-3868 (1996).

\bibitem{PAW}
P. E. Bl$\ddot{o}$chl, Phys. Rev. B. {\bf 50}, 17953 (1994);
G. Kresse and J. Joubert, Phys. Rev. B. {\bf 59}, 1758 (1999).
%
\bibitem{VASP}
G. Kress and J. Hafner, Phys. Rev. B. {\bf 48}, 13115 (1993);
G. Kress and J. Furthm$\ddot{u}$ller, Comput. Mater. Sci. {\bf 6}, 15 (1996); Phys. Rev. B. {\bf 54}, 11169 (1996).

\bibitem{HSE} A. V. Krukau, O. A. Vydrov, A. F. Izmaylov, and G. E. Scuseria, J. Chem. Phys. \textbf{125}, 224106
(2006)

\bibitem{sm}
Supplementary Material

%
%

\bibitem{bhz}
B.A. Bernevig, T.L. Hughes, and S-C. Zhang. Science 314, 1757 (2006).

\bibitem{kariyado1}
T. Kariyado and M. Ogata. J. Phys. Soc. Jpn. 80 083704 (2011).

\bibitem{oxide}
J. Chakhalian,  A. J. Millis, and J. Rondinelli. Nature Materials 11, 92Ð94 (2012).

\bibitem{nagaosa}
D. Xiao, W. Zhu, Y. Ran, N. Nagaosa	, and S. Okamoto. Nat. Comm. 2:596 (2011).

\bibitem{oxide2}
A. Ruegg, C. Mitra, A. A. Demkov, and G.A. Fiete. Phys. Rev. B 85, 245131 (2012).

\bibitem{oxide3}
D. Doennig, W.E. Pickett, and R. Pentcheva. Phys. Rev. B 89, 121110(R) (2014).

\bibitem{elder}
W.J. Elder, R.M. Ward, and J. Zhang. Phys. Rev. B {\bf 83}, 165210 (2011).

\bibitem{luttinger}
J.M. Luttinger, Phys. Rev. {\bf 102}, 1030 (1956).

\bibitem{zunger}
J. Vidal, X. Zhang, L. Yu, J.-W. Luo, and A. Zunger. Phys. Rev. B 84, 041109(R) (2011).

\end{thebibliography}
\end{document}